\documentclass[prb,twocolumn,preprintnumbers,amsmath,amssymb]{revtex4}

\usepackage{graphicx}% Include  files
\usepackage{dcolumn}% Align table columns on decimal point
\usepackage{bm}% bold math
\usepackage{epstopdf}
\begin{document}

\title{A molecular dynamics study of chemical gelation in a patchy particle model}

\author{Silvia Corezzi,* $^{a b}$ Cristiano De Michele,$^{c d}$ Emanuela Zaccarelli,$^{c d}$
Daniele Fioretto$^{b c}$ and Francesco Sciortino$^{c d}$}
    \affiliation{$^a$CNR--INFM Polylab, Universit$\grave{a}$ di Pisa, Largo
Pontercorvo 3, I-56127 Pisa, Italy\\
$^b$Dipartimento di Fisica, Universit$\grave{a}$ di Perugia, Via
A. Pascoli, I-06100 Perugia, Italy\\
$^c$CNR--INFM CRS Soft, Universit$\grave{a}$ di Roma ``La
Sapienza", P. A. Moro 2, I-00185 Roma, Italy\\
$^d$Dipartimento di Fisica, Universit$\grave{a}$ di Roma ``La
Sapienza", P. A. Moro 2, I-00185 Roma, Italy}

\begin{abstract}
We report event--driven molecular dynamics simulations of the
irreversible gelation of hard ellipsoids of revolution containing
several associating groups, characterizing how the cluster size
distribution evolves as a function of the extent of reaction,
both below and above the gel point. We find that in a very large
interval of values of the extent of reaction, parameter--free
mean--field predictions are extremely accurate, providing
evidence that in this model the Ginzburg zone near the gel point,
where non--mean field effects are important, is very limited. We
also find that the Flory's hypothesis for the post--gelation
regime properly describes the connectivity of the clusters even
if the long--time limit of the extent of reaction does not reach
the fully reacted state. This study shows that irreversibly
aggregating asymmetric hard--core patchy particles may provide a
close realization of the mean--field model, for which available
theoretical predictions may help control the structure and the
connectivity of the gel state. Besides chemical gels, the model
is relevant to network--forming soft materials like systems with
bioselective interactions, functionalized molecules and patchy
colloids.

\end{abstract}

\date{\today}
\maketitle

\section{Introduction}

\begin{figure}[t]
\includegraphics[width=7. cm]{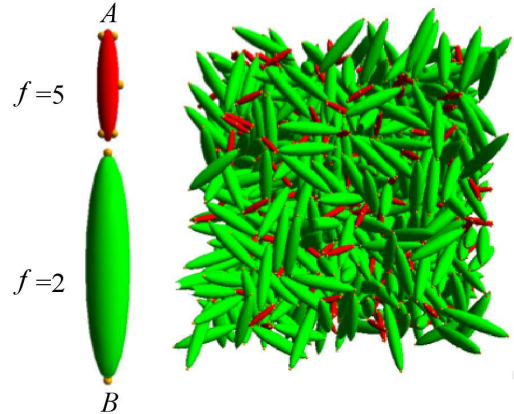}
\caption{\label{fig:particle} Graphic description of the $A$ and
$B$ particles (left) and snapshot of the simulated system
(right). The centers of the small spheres locate the bonding sites
on the surface of the hard--core particle.}
\end{figure}

Irreversible polymerization is a mechanism of self--organization
of molecules which proceeds via the formation of covalent bonds
between pairs of mutually--reactive groups.\cite{FloryBOOK,
ColbyBOOK, burchard} If monomers with functionality (number $f$
of reactive groups on a monomer) greater than two are present,
branched molecules grow by reactions and convert the system from
a fluid of monomers into a well connected cross--linked network,
giving rise to a chemical gelation process. At the gel point, a
persistent network spanning the sample first appears; the system
is then prevented from flowing, yet not arrested on a mesoscopic
length scale. The development of a network structure results, for
example, from step polymerization, chain addition polymerization
and cross--linking of polymer chains.\cite{YoungLovell,
Thermosets} The same phenomenon is also observed in colloids and
other soft materials when the thermodynamics and the molecular
architecture favor the formation of a limited number of strong
interactions (i.e., with attraction strength much larger than the
thermal energy) between different particles. Chemical gelation
has been extensively studied in the past, starting from the
pioneering work of Flory and Stockmayer\cite{FloryBOOK,
StockmayerJPS1952} who developed the first mean--field
description of gelation, providing expressions for the cluster
size distribution as a function of the extent of reaction and the
{\it critical} behavior of the connectivity properties close to
gelation. More appropriate descriptions based on geometric
percolation concepts have, in the late seventies, focused on the
non--mean field character of the transition, which reveals itself
near the gel point, extending to percolation the ideas developed
in the study of the properties of systems close to a
second--order critical point. Several important numerical
studies,\cite{StaufferJCSFT1976, Manneville1981, HerrmannPRL1982,
Pandey, ClercAP1983, BansilMACRO1984, LeungJCP1984, GuptaJCP1991,
LairezJPF1991, Gimel, VernonPRE2001, DelGadoPRE2002}
---most of them based on simulations on lattice
---have focused on the critical behavior close to the
percolation point, providing evidence of the percolative nature
of the transition and accurate estimates of the percolation
critical exponents. As in critical phenomena, a crossover from
mean--field to percolation behavior is expected close to the gel
transition.\cite{ginzburg} But, how the microscopic properties of
the system control the location of the crossover (i.e., how wide
is the region where the mean--field description applies) and how
accurate is the mean--field description far from the percolation
point is not completely understood. Another important open
question regards the connectivity properties of chemical gels
well beyond percolation.\cite{rubinstein} Even in the mean--field
approximation, several possibilities for the post--gel solutions
have been proposed, based on different assumptions on the
reactivity of sites located on the infinite
cluster.\cite{VanDongenJSP1997, rubinstein} Different propositions
predict different cluster--size distributions above the gel point
and a different evolution with time for the extent of reaction.

Here we introduce a model inspired by stepwise polymerization of
bifunctional diglycidyl--ether of \mbox{bisphenol--A} ($B$
particles in the following) with pentafunctional
diethylenetriamine ($A$ particles).\cite{CorezziPRL2005} To
incorporate excluded volume and shape effects, each type of
molecule is represented as hard homogeneous ellipsoid of
appropriate length, whose surface is decorated in a predefined
geometry by $f$ identical reactive sites per particle (see
Figure~\ref{fig:particle}). In this respect, the model is also
representative of colloidal particles functionalized with a
limited number of patchy attractive sites,\cite{Glotz-Solomon}
where the selectivity of the interaction is often achieved
building on biological specificity.\cite{hiddessen, DNA, DNAsoft}
The off--lattice evolution of the system is studied via
event--driven molecular dynamics simulations, using a novel code
which specifically extends to ellipsoidal particles the algorithm
previously designed for patchy spheres.\cite{pwm} Differently
from previous studies, we do not focus on the critical properties
close to the gel--point but study in detail the development of the
irreversible gelation process and the properties of the cluster
size distribution in the pre-- and post--gelation regime.

We find that the dynamic evolution of the system produces an
irreversible (chemical) gelation process whose connectivity
properties can be described, in a very large window of the extent
of reaction, with the Flory--Stockmayer (FS)
predictions.\cite{FloryBOOK, ColbyBOOK, StockmayerJPS1952} This
offers to us the possibility to address, in a well controlled
model, the kinetics of the aggregation and to evaluate the extent
of reaction at which the breakdown of the Flory post--gel
solution takes place.

\section{Method}

We study a 5:2 binary mixture composed of $N_A=480$ ellipsoids of
type $A$ and $N_{B}=1200$ ellipsoids of type $B$, for a total of
$N=1680$ particles. $A$ particles are modeled as hard ellipsoids
of revolution with axes $a=b=2\sigma$ and $c=10\sigma$ and mass
$m$; $B$ particles have axes $a=b=4\sigma$ and $c=20\sigma$, mass
$3.4 m$. Simulations are performed at a fixed packing fraction
$\phi=0.3$. Five (two) sites are rigidly anchored on the surface
of the $A$ ($B$) particles, as described in
Fig.~\ref{fig:particle}. Sites on $A$ particles can only react
with sites on $B$ particles. Every time, during the dynamic
evolution, the distance between two mutually--reactive sites
becomes smaller than a predefined distance $\delta=0.2 \sigma$, a
new bond is formed between the particles. To model irreversible
gelation, once a bond is formed, it is made irreversible by
switching on an infinite barrier at distance $r^{ij}_{AB}=\delta$
between the sites $i$ and $j$ involved, which prevents both the
formation of new bonds in the same sites and the breaking of the
existing one. Hence, the newly formed bond cannot break any
longer, and the maximum distance between the two reacted sites is
constrained to remain smaller than $\delta$. Similarly, the two
reacted sites cannot form further bonds with available unreacted
sites. The composition of the system and the particle
functionality are such that the reactive sites of type $A$ and
$B$ are initially present in equal number,
$f_{A}N_{A}=f_{B}N_{B}$, which in principle allows the formation
of a fully bonded state in which all the sites have reacted. This
offers a way to properly define the extent of reaction as the
ratio $p$ between the number of bonds present in a configuration
and the maximum number of possible bonds $f_{A}N_{A}$.

Between bond--formation events, the system propagates according to
Newtonian dynamics at temperature $T=1.0$. As in standard
event--driven codes, the configuration of the system is propagated
from one collisional event to the next one. Note that temperature
only controls the time scale of exploration of space, by
modulating the average particle's velocity. An average over 40
independent starting configurations is performed to improve
statistics.

\begin{figure}
\includegraphics[width=10 cm]{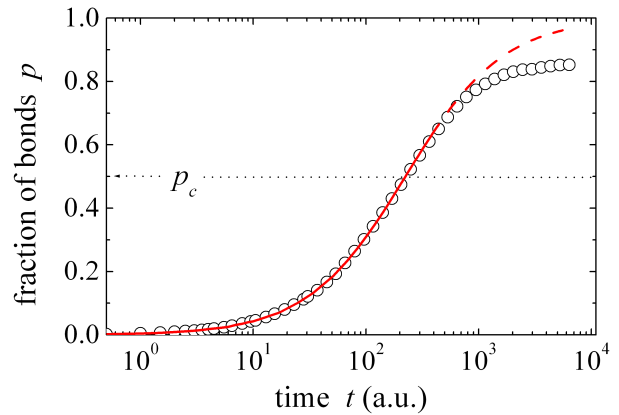}
\vspace{-0.8 cm} \caption{\label{fig:conv} Time dependence of the
fraction of bonds $p$. Symbols: simulation results (averaged over
40 independent realizations). For the chosen stoichiometry, $p$
coincides with the reacted fraction of $A$ reactive sites, i.e.
the $A$ conversion, or equivalently with the reacted fraction of
$B$ sites, i.e. the $B$ conversion. $p=1$ would indicate that all
possible bonding sites have reacted. Time is measured in
arbitrary units. Line: $p(t)=kt/(1+kt)$, with the fit--parameter
$k$ fixing the time scale. This functional form is expected when
any pair of reactive groups in the system is allowed to react,
but loops do not occur in finite size
clusters.\cite{VanDongenJSP1997}}
\end{figure}

\section{Results}

In the starting configurations no bonds are present by
construction. As a function of time, the fraction $p$ of formed
bonds ---a measure of the state of advancement of the reaction---
increases monotonically, until most of the particles are
connected in one single cluster (Figure~\ref{fig:conv}). As a
result, $p$ saturates around $0.86$, despite the fact that an
equal number of reactive sites of type $A$ and $B$ is initially
present in the system.

Flory and Stockmayer\cite{FloryBOOK, StockmayerJPS1952} laid out
the basic relations between extent of reaction and resulting
structure in step polymerizations, on the assumptions that all
functional groups of a given type are equally reactive, all
groups react independently of one another, and that ring
formation does not occur in molecular species of finite size.
Only when $p$ exceeds a critical value $p_{c}$ infinitely large
molecules can grow.\cite{FloryBOOK} In this respect the FS theory
describes the gelation transition as the random percolation of
permanent bonds on a loopless lattice.\cite{Stauffer1992} The
present model satisfies the conditions of equal and independent
reactivity of all reactive sites. The absence of closed bonding
loops in finite size clusters is not a priori implemented; as we
will show in the following, however, such a condition ---favored
by the poor flexibility of the bonded particles and their
elongated shape, the absence of an underlying lattice and the
asymmetric location of the reactive sites--- is valid in a
surprisingly wide region of $p$ values.

\begin{figure}
\includegraphics[width=7.5 cm]{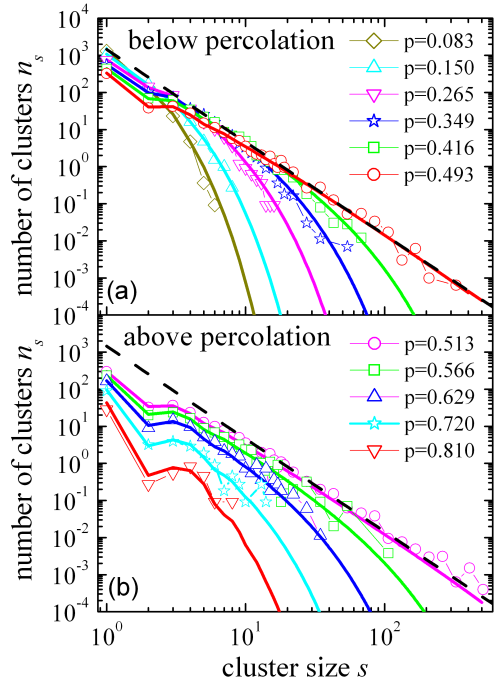}
\caption{\label{fig:distr} Distribution of finite size clusters
$n_{s}$ for different fraction of bonds $p$ (a) below and (b)
above percolation. Points are simulation data and lines are the
corresponding theoretical curves according to FS. The
dashed line represents the power law $n_{s}\sim s^{-2.5}$.}
\end{figure}

The FS theory predicts the $p$ dependence of the cluster size
distribution in the very general case of a mixture of monomers
bearing mutually reactive groups.\cite{StockmayerJPS1952} In the
present case, the number $n_{lm}$ of clusters containing $l$
bifunctional particles and $m$ pentafunctional ones can be
written as
\begin{eqnarray}
n_{lm}=N_{B} N_{A}  p^{l+m-1}(1-p)^{3m+2}
w_{lm}\label{eq:nlm}\\
w_{lm}=\frac{(4m)!} {(l-m+1)!(4m-l+1)!m!}\nonumber
\end{eqnarray}
and the number of clusters of size $s$ is obtained by summing
over all contributions such that $l+m=s$, i.e., $n_{s}=\sum
_{lm,l+m=s}n_{lm}$. As shown in Figure~\ref{fig:distr}a on
increasing $p$, the $n_{s}$ distribution becomes broader and
broader and develops a power--law tail. The theory predicts a
gelation transition when
$p_c=1/\sqrt{(f_A-1)(f_B-1)}=0.5$.\cite{FloryBOOK,
StockmayerJPS1952} Even close to $p=0.5$, the FS prediction
---which conforms to the prediction of random percolation on a
Bethe (loopless) lattice where $n_{s}\sim s^{-2.5}$ at the
percolation threshold--- is consistent with the numerical data.
On further increasing $p$ (Figure~\ref{fig:distr}b), the
distribution of finite size clusters progressively shrinks, and
only small clusters survive. Data show that Eq~\ref{eq:nlm}, with
no fitting parameters, predicts rather well the numerical
distribution at any extent of polymerization, both below and
above the point where the system is expected to percolate,
including details such that the local minimum at $s=2$.

\begin{figure}
\includegraphics[width=7.5 cm]{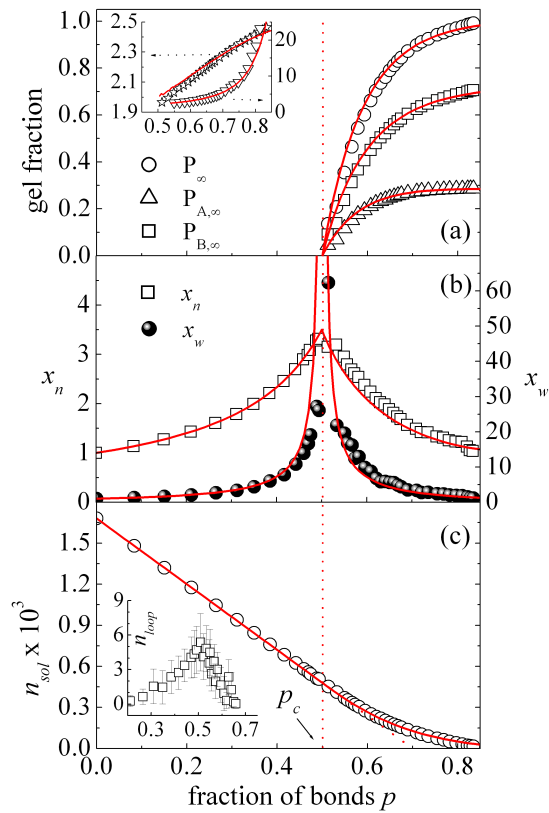}
\caption{\label{fig:GELsol} (a) Gel fraction $P_{\infty}$ and its
partition between particles of type $A$ ($P_{A,\infty}$) and $B$
($P_{B,\infty}$) vs the fraction of bonds $p$ (i.e. the extent of
reaction in the system). The inset shows the proportion of $B$
particles to $A$ particles in gel ($N_{B,\infty}/N_{A,\infty}$
--- left axis) and in sol ($N_{B,sol}/N_{A,sol}$
--- right axis) vs $p$. (b) Number and weight average cluster
size ($x_{n}$ and $x_{w}$) prior to gelation and for the sol
after gelation vs the fraction of bonds $p$. (c) Relation between
the number of finite size clusters (molecules in the sol)
${n}_{sol}$ and the fraction of bonds $p$. The inset shows the
number of loops ${n}_{loop}$ vs $p$. In all panels, symbols are
simulation results and solid lines FS predictions.}
\end{figure}

To compare with the mean--field prediction of gelation at
$p_{c}=0.5$, we examine the connectivity properties of the
aggregates for each studied value of $p$, searching for the
presence of clusters which are infinite under periodic boundary
conditions. We find that configurations at $p=0.497\pm 0.008$
have not yet developed a percolating structure while
configurations at $p=0.513\pm 0.007$ have. Hence, we locate the
gel point at $p_{c}=0.505\pm 0.007$, in close agreement with the
theoretical mean--field expectations. Beyond this point, the
material which belongs to the infinite (percolating) network
$N_{\infty}$ constitutes the \emph{gel}, while the soluble
material formed by the finite clusters which remain interspersed
within the giant network constitutes the \emph{sol}.
Figure~\ref{fig:GELsol}a shows that the fraction of gel
$P_{\infty}=N_{\infty}/N$ and even its partition between
particles of type $A$ ($P_{A,\infty}=N_{A,\infty}/N$) and $B$
($P_{B,\infty}=N_{B,\infty}/N$) calculated according to the FS
theory,\cite{MillerMACRO1976} properly represent the simulation
results throughout the polymerization process. Indeed, the
proportion of $B$ particles to $A$ particles in gel and in sol is
a function of $p$ (see inset). The relative amount of $B$
particles in the sol ($N_{B,sol}/N_{A,sol}$) increases as a
consequence of the preferential transfer of the $A$ particles
(having more reactive sites) to the gel, in a way that the
fraction $p_{sol}$ of sites $B$ in the sol that have reacted
(extent of reaction in the sol) differs from the total fraction
$p$ of sites $B$ reacted (extent of reaction in the system). The
constitution of the sol (Figure~\ref{fig:distr}(b)) results to be
the same as that of a smaller system made of $N_{A,sol}$
particles of type $A$ and $N_{B,sol}$ particles of type $B$
reacted up to the extent $p_{sol}$.\cite{FloryBOOK, MillerPES1979}

The evolution of the cluster size distribution can be quantified
by the number ($x_n$) and weight average ($x_w$) cluster sizes of
the sol, defined as $x_{n}=\sum_{s}sn_{s}/\sum_{s}n_{s}$ and
$x_{w}=\sum_{s}s^{2}n_{s}/\sum_{s}sn_{s}$. The numerical results
and the FS theoretical predictions are shown in
Figure~\ref{fig:GELsol}b. Both averages increase before gelation;
then, they regress in the sol existing beyond the gel point,
since large clusters are preferentially incorporated into the gel
network. While $x_{n}$ increases only slightly up to the gel
point, never exceeding 3.5, $x_{w}$ increases sharply in
proximity of $p_{c}$ as well as sharply decreases beyond this
point, consistently with the fact that $x_w$ is singular at
percolation being dominated by large clusters. Again, simulation
data agree very well with FS predictions. Discrepancies between
theory and simulation ---which reveal the mean--field character of
the FS theory--- only concern the range of $p$ very near $p_{c}$,
suggesting that for this model the crossover from mean--field to
percolation is very close to the gel point --- i.e., the Ginzburg
zone\cite{ginzburg} near the gel point, where non--mean field
effects are important, is very limited. A finite--size study very
close to the critical point would be requested to accurately
locate the percolation point and the critical exponents, a
calculation beyond the scope of the present work.

From a physical point of view, the change from mean--field to
percolation universality class is rooted in the presence of
bonding loops in the clusters of finite size, which pre-empts the
possibility to predict the cluster size distribution. The
realistic estimate of the percolation threshold and the agreement
between theory and simulation (Fig.~\ref{fig:distr}) suggest that
the present model strongly disfavors the formation of loops in
finite clusters, at least for cluster sizes probed in this
finite--size system. As a test, we evaluate the total number of
finite (sol) clusters ${n}_{sol}=\sum_{s}n_{s}$ as a function of
the extent of reaction. If finite clusters do not contain closed
loops, ${n}_{sol}$ equals the number of particles in the sol
minus the number of bonds, since each added bond decreases the
number of clusters by one. This applies equally to the system
preceding gelation, or to the sol existing beyond the gel point.
Thus, at $p<p_{c}$ (pre--gelation) the relation between
${n}_{sol}$ and $p$ is linear, i.e. ${n}_{sol}=N - 2N_{B}p$. At
$p>p_{c}$ (post--gelation), ${n}_{sol}$ can be calculated as
${n}_{sol}= N_{sol} - 2N_{B,sol}p_{sol}$, where $N_{sol}$ is the
number of particles in the sol fraction ($N_{B,sol}$ of which
bear reactive sites of type $B$), and $p_{sol}\neq p$ is the
reacted fraction of sites B in the sol. Hence, the relation
between ${n}_{sol}$ and $p$ crosses to a nonlinear behavior, so
that the number of clusters becomes one when $p=1$. As shown in
Figure~\ref{fig:GELsol}c, the number of finite clusters found in
the simulation data conforms to the theoretical expectation for
all $p$ values, both below and above the gel point. Hence, as a
first approximation, loops are only present in the infinite
(percolating) cluster and do not significantly alter the
distribution of the finite size clusters, both below and above
percolation. The difference between ${n}_{sol}$ found in
simulation and the value predicted by the FS theory counts the
number of loops in the sol, ${n}_{loop}$. Such a quantity is
shown in the inset of Figure~\ref{fig:GELsol}c. The maximum value
of ${n}_{loop}$, achieved for $p \sim p_{c}$, corresponds to
$0.2\%$ of the total number of bonds. This demonstrates that
intramolecular bonds within finite clusters can be neglected,
consistent with the Flory hypothesis for the post--gelation
regime\cite{rubinstein}. Figure~\ref{fig:GELsol}c also shows that
the linear relation between ${n}_{sol}$ and $p$ is valid also
after the gel point (up to $p \approx 0.6$). This finding is in
full agreement with recent experimental
studies\cite{CorezziPRL2005, CorezziJPCM2005, VolponiMACRO2007}
on the polymerization of bifunctional diglycidyl--ether of
\mbox{bisphenol--A} with pentafunctional diethylenetriamine, also
suggesting that the number of cyclic connections in the infinite
cluster is negligible well above $p_{c}$.

As a further confirmation of the absence of closed loops we
compare the time evolution of $p$ with the prediction of the
mean--field kinetic modeling of polymerization, based on the
solution of the Smoluchowski coagulation
equation.\cite{ZiffJSP1980, GalinaAPS1998} For loopless
aggregation, $p(t)$ is predicted to follow
\begin{equation}
p(t)=\frac{kt}{1+kt},
\end{equation}
where the fit--parameter $k$, which has the meaning of a bond
kinetic constant, fixes the time scale of the aggregation process.
The time evolution of $p$ is found to perfectly agree with the
theoretical predictions\cite{VanDongenJSP1997} (see
Figure~\ref{fig:conv}) up to $p \approx  0.6$, i.e. beyond $p_c$.
While the prediction would suggest that $p(t \rightarrow
\infty)=1$ (dash line in Figure~\ref{fig:conv}), the simulation
shows that the formation of a percolating structure prevents the
possibility of completing the chemical reaction, leaving a finite
number of unreacted sites frozen in the structure. As shown above
(Figure~\ref{fig:distr}), even in this frozen state the cluster
size distribution is provided by the Flory's post--gel
hypothesis. Such a feature is not captured by the mean--field
Smoluchowski equation in which spatial information in the kernels
are neglected.

\section{Conclusions}

A binary mixture of patchy hard ellipsoids undergoing chemical
gelation displays a very large interval of the extent of reaction
in which parameter--free mean--field predictions are extremely
accurate. The connectivity properties of the model are properly
described --- without any fitting parameter --- both below and
above percolation by the mean--field loopless classical FS
theory.\cite{FloryBOOK, VanDongenJSP1997} The mean--field cluster
size distribution for the sol component is found to be valid for
all values of the extent of reaction, both below and above the
gel point, suggesting that for the present model, the Flory's
hypothesis for the post--gelation regime properly describes the
irreversible aggregation phenomenon, despite the explicit
consideration of the excluded volume.

The absence of loops in finite size clusters, which is not
assumed by the model, results from the specific geometry of the
bonding pattern and by the presence of the excluded volume
interactions, disfavoring the formation of ordered bonding
domains. Hence, the geometry of the particles and the location of
the reactive sites on them may play a significant role in the
stabilization of the mean--field universality class with respect
to the percolation universality class,\cite{StaufferAPS1982}
locating the crossover between the two classes\cite{ginzburg}
very close to the gel point. The present study shows that
irreversibly aggregating asymmetric hard--core patchy particles,
even if excluded volume effects are properly taken into account,
may provide a close realization of the FS predictions in a wide
range of $p$ values. The model thus offers a starting point ---for
which theoretical predictions are available--- for further
investigations of the gelation process and for a more precise
control over the structure and connectivity of the gel state. In
particular, a full and detailed structural information can be
known along with the dynamics of the system, which is potentially
useful to investigate the relation between structural
heterogeneity and heterogeneous dynamics,\cite{VolponiMACRO2007}
and to shed light on the microscopic aspects of the dynamic
crossover from short\cite{CorezziPRL2006} to long relaxation
times,\cite{CorezziNATURE2002} during irreversible polymerization.

While the structural properties are all well--described by the FS
theory, the evolution of the extent of reaction, modeled via the
coagulation Smoluchowski equation, is properly described by the
theory only in the pre--gelation region. After gelation, kinetic
constraints due to the absence of mobility of the reactive sites
anchored to the percolating cluster or to smaller clusters
trapped inside the percolating matrix prevent the completion of
the reaction and the extent of reaction freezes (to $p\approx
0.86$ in the present case) before reaching one (as Eq.~2 would
predict). A proper modeling of the long--time behavior will
require the insertion of spatial information inside the kernels
entering the Smoluchowski equation. The freezing of the extent of
reaction at long times correspondingly freezes the cluster size
distribution to that predicted by Flory for the reached $p$ value.

In the present model, the entire polymerization process proceeds
via a sequence of FS cluster size distributions, determined by
$p(t)$. Recently, it has been shown that the FS theory properly
describes also equilibrium clustering in patchy particle systems
when $p$ is a function of temperature and
density.\cite{BianchiJPCB2007} It is thus tempting to speculate
that for loopless models, irreversible evolution can be put in
correspondence with a sequence of \textit{equilibrium} states
which could be sampled in the same system for finite values of
the ratio between temperature and bonding depth. If this is
indeed the case, chemical gelation could be formally described as
a deep quench limit of physical gelation. This correspondence
would facilitate the transfer of knowledge from recent studies of
equilibrium gels\cite{BianchiPRL2007, ZaccarelliREVIEW2007} to
chemical ones. Concepts developed for irreversible aggregation of
colloidal particles, like diffusion-- and reaction--limited
cluster--cluster aggregation, could be connected to chemical
gelation. Work in this direction is ongoing.

We acknowledge support from MIUR-PRIN. We thank P. Tartaglia for interesting discussions.

\end{document}